\documentclass[11pt]{article}
\usepackage{amsmath}
\usepackage{amssymb}
\usepackage{graphics}

\def\cp{\mathbb{CP}^2}
\newcommand\beq{\begin{eqnarray}}
\newcommand\eeq{\end{eqnarray}}

\begin{document}

\begin{flushright}
CCTP-2010-11
\end{flushright}
%\smallskip

\begin{center}
\LARGE Properties of Schr\"odinger Black Holes from AdS Space \\
\large
\bigskip\bigskip
Bom~Soo~Kim$^{a,b}$ and Daisuke Yamada$^{a}$
        \bigskip
        \\ $^a$
       {\normalsize {\it Crete Center for Theoretical Physics},
       {\it University of Crete},             \\
        {\it P.O. Box 2208, 71003 Heraklion, Crete, GREECE }}
        \\ $^b$ 
        {\normalsize {\it IESL, Foundation for Research and Technology - Hellas, }\\
        {\it P.O. Box 1527, 71110 Heraklion, Crete, GREECE }}
          \bigskip        \\
        {\normalsize  {\tt bskim,dyamada@physics.uoc.gr} } 
\end{center}      \smallskip

%%%%%%%%%%%%%%%%%%%%%%%%%%%%%%%%%%%%%%%%%%%%%%%%%%%%%%%%%%%%%%%%%%%%%
\begin{abstract}
%%%%%%%%%%%%%%%%%%%%%%%%%%%%%%%%%%%%%%%%%%%%%%%%%%%%%%%%%%%%%%%%%%%%%
Properties of Schr\"odinger black holes are derived from those of 
AdS black holes expressed in light-cone coordinates with a
particular normalization. 
Unlike the usual construction from an AdS black hole using a null Melvin twist, 
an AdS black hole in light-cone is simple and has a well-defined
Brown-York procedure with the standard counterterms. 
Our procedure is demonstrated by 
the computation of the DC conductivity and the derivation
of the R-charged black hole thermodynamic properties.

\end{abstract}

\bigskip

%%%%%%%%%%%%%%%%%%%%%%%%%%%%%%%%%%%%%%%%%%%%%%%%%%%%%%%%%%%%%%%%%%%%%
\tableofcontents
%%%%%%%%%%%%%%%%%%%%%%%%%%%%%%%%%%%%%%%%%%%%%%%%%%%%%%%%%%%%%%%%%%%%%

%\pagebreak

%%%%%%%%%%%%%%%%%%%%%%%%%%%%%%%%%%%%%%%%%%%%%%%%%%%%%%%%%%%%%%%%%%%%%
\section{Introduction}\label{sec:intro}
%%%%%%%%%%%%%%%%%%%%%%%%%%%%%%%%%%%%%%%%%%%%%%%%%%%%%%%%%%%%%%%%%%%%%

The exploration of non-relativistic generalizations of the AdS/CFT correspondence
\cite{Maldacena:1997re,Gubser:1998bc,Witten:1998qj}, 
initiated by Son~\cite{Son:2008ye} and by
Balasubramanian and McGreevy~\cite{Balasubramanian:2008dm}, 
has become an active research area.
The original works focused on a geometric realization of the non-relativistic
conformal symmetry, also known as Schr\"odinger symmetry, and the proposed
geometry (``Schr\"odinger space'') successfully and naturally fitted into
the usual scheme of the AdS/CFT correspondence.
Soon after, the system was generalized to non-zero temperature by incorporating
black holes in the geometry \cite{Herzog:2008wg,Maldacena:2008wh,Adams:2008wt}.
In addition, it was realized that the five
dimensional Schr\"odinger space can be obtained from the AdS$_5\times S^5$
solution of type IIB supergravity by applying a series of transformations
known as a null Melvin twist \cite{Alishahiha:2003ru,Gimon:2003xk}.
Subsequently, the analysis of Schr\"odinger space, especially through the null
Melvin twist, has become mainstream as a geometric realization of
Schr\"odinger symmetry.

However, before the finite temperature generalization was considered,
it was found by Goldberger~\cite{Goldberger:2008vg} and by
Barbon and Fuertes~\cite{Barbon:2008bg} that the Schr\"odinger
symmetry can be geometrically realized in the pure AdS space, without
a deformation. Their procedure is very similar to the light-front quantization 
of relativistic field theories, which can be reduced to Galilean invariant 
non-relativistic {\it analogues} \cite{Susskind:1967rg,Bardakci:1969dv,Chang:1968bh}.%
\footnote{
	The purpose of the infinite momentum frame is to investigate the question : 
	"Can one write Schr\"odinger (Galilean-invariant) theories at infinite momentum 
	that completely describe interacting relativistic system?" In particular, 
	a theory (at infinite momentum) demands Poincar\'e invariance, unitarity, and positivity 
	of mass spectrum \cite{Bardakci:1969dv}.
	}
The key difference, however, is that the authors of 
References~\cite{Goldberger:2008vg,Barbon:2008bg} 
project the theory onto a fixed momentum in one of the light-cone
directions and identify the other light-cone coordinate as the time of the resulting 
non-relativistic theory.  
These two key procedures differentiate the system from those of the infinite momentum frame, 
in which the momentum projection were not done.
Moreover, light-cone directions do not have the interpretation as time and
space, even though one of them is sometimes called ``light-cone time.''
It is just a convenient frame to work for some cases, and in fact, the
formalism is Lorentz invariant. 

In addition to these two important operations, we 
use the light-cone coordinates with a particular normalization \cite{Maldacena:2008wh}
\begin{align}\label{eq:introLC}
  x^+ = b (t+x)
  \;,\quad
  x^- = \frac{1}{2b} (t-x)
  \;,
\end{align} 
even for the zero temperature case.
This transformation can be thought as a two-step procedure: 
a boost in the $x$-direction with rapidity $\log b$  
followed by going to light-cone coordinates. 
To ensure a definite dynamical exponent z=2, 
we assign $[b]$ (a scaling dimension in the unit of mass) as $-1$, 
and thus $[x^+] = -2$ and $[x^-] = 0$. $x^-$ is invariant under the scaling transformation, 
which is crucial for the system to have 
special conformal invariance in its symmetry group.  
The parameter $b$ exactly matches the extra parameter generated in the null Melvin twist.
Incidentally, the light-cone coordinates (\ref{eq:introLC}) have been adopted in the previous
works with the null Melvin twist 
because it is the unique normalization that makes the metric of
Schr\"odinger space independent of the parameter $b$ in its asymptotic boundary.
But it has not been emphasized that this is originated from a boost parameter. 
%\footnote{
%  Even though the parameter $b$ is originated from a boost parameter, it can not be scaled away 
%  due to its non-trivial scaling dimension as well as the momentum projection of 
%  the $x^-$ direction even in the zero temperature case or 
%  the asymptotic geometry of Schr\"odinger space.
%}
Thus the resulting physical quantities, described by the light-cone coordinates with 
modified scaling dimensions $[x^+]=-2$ and $[x^-]=0$, are different from those of the 
original AdS space.   

Therefore, assigning different scaling dimensions to the light-cone coordinates, 
the momentum projection and the re-interpretation of time completely separate the system from
the relativistically causal theory. One should not carry over
relativistic reasoning, such as causal structure from  the original AdS spacetime.
This implies that pure AdS space in light-cone coordinates with the three procedures 
can rightly be a conceivable dual of a non-relativistic field theory. The technique adopted in
References~\cite{Goldberger:2008vg,Barbon:2008bg,Maldacena:2008wh}, in a sense,
can be regarded as the AdS/CFT gravitational counterpart to
the field theory in the light-front, but the system is truly non-relativistic,
rather than a non-relativistic analogue, 
as a result of the three key procedures mentioned above.

Despite interesting aspects and its sheer simplicity,
the approach of Goldberger, Barbon and Fuertes has become less popular
and non-zero temperature generalizations along this line have not been
pursued explicitly.
One exception is Reference~\cite{Maldacena:2008wh} where
it is briefly mentioned that
the thermodynamic quantities of a planar Schr\"odinger black hole can
be obtained also from a planar AdS black hole
expressed in the light-cone coordinates with a particular normalization
given in equation (\ref{eq:introLC}).% 
\footnote{
 This observation was adopted and the result was used
 in the discussion of the hydrodynamics from the Schr\"odinger
 black holes in Reference~\cite{Rangamani:2008gi}.
}
%\begin{align}\label{eq:introLC2}
%  x^+ = b (t+x)
%  \;,\quad
%  x^- = \frac{1}{2b} (t-x)
%  \;.
%\end{align}

In this paper, we pursue the direction of
References~\cite{Goldberger:2008vg,Barbon:2008bg,Maldacena:2008wh}. 
We explicitly demonstrate that the introduction of the 
light-cone coordinates (\ref{eq:introLC}) to a relativistic AdS system of interest, 
with the three key procedures, 
is enough to reproduce the thermodynamic and transport properties
of the Schr\"odinger counterpart.
The advantages of this approach are the following.
Most importantly, the well-defined nature of the boundary permits
the computation of thermodynamic quantities through the Brown-York
procedure \cite{Brown:1992br} with the standard counterterms of
Balasubramanian and Kraus \cite{Balasubramanian:1999re}.
This is carefully laid out in Section~\ref{sec:procedure} by taking
the planar AdS$_5$ black hole as a prototype example.
The Brown-York procedure completely fails when applied to the Schr\"odinger
black holes --- due to the unusual boundary structure%
\footnote{See \cite{Horava:2009vy} for the earlier attempt to resolve the puzzles 
	of the holographic renormalization for the spacetime with unusual boundary structure  
	including the Schr\"odinger and Lifshitz spacetime 
	using anisotropic scaling.
	}
 --- and the derivations of the thermodynamic quantities
have been very awkward and {\it ad hoc}.
In particular, all the derivations effectively assume the first law
of thermodynamics and the independent check of the first law for the
derived quantities has not been possible.
(See, however, the work of Ross and Saremi \cite{Ross:2009ar} in which
the thermodynamic quantities are successfully derived by adopting modified
definition of the stress tensor.)

It is worth while to mention here that there are two technical 
modifications in computing conserved quantities along the two light-cone 
coordinates. This is required due to the cross term $dx ^+ dx^-$ in AdS in light-cone compared 
to its original AdS space. We systematically explain them in equations 
(\ref{DefMass}), (\ref{DefDensity}) and (\ref{DefMomentum}) of 
Section~\ref{sec:procedure} following carefully \cite{Brown:1992br}. 
Identifying this modifications is 
possible because of our clear procedure with a well defined boundary structure.

Another important advantage of our construction is its simplicity.
Generating an asymptotically Schr\"odinger spacetime
through the null Melvin twist is a fairly complicated procedure.
This is especially so when the original metric has non-vanishing off-diagonal
components in time and space directions, and the resulting RR-potentials
becomes even more complicated.
One such intricate example is an R-charged black hole.
However, the procedure that we adopt is no more complicated than usual
relativistic calculations, and in Section~\ref{sec:RCharged},
thermodynamic quantities of the Schr\"odinger
R-charged black hole are derived for the planar, spherical and hyperbolic
cases including the finite Liu-Sabra counter terms \cite{Liu:2004it} for 
the first law of thermodynamics to be satisfied.
In that section, we also discuss phase diagrams of the black holes
in the planar and spherical cases.

In Section~\ref{sec:DC}, we demonstrate that our method is not only
effective in deriving thermodynamic quantities but also 
Minkowski two-point correlators and conductivities.
In particular, we derive the non-relativistic counterpart of
the DC conductivity computed by Karch and O'Bannon \cite{Karch:2007pd}.
This result is compared to the conductivity calculated directly from
the Schr\"odinger space by Ammon {\it et al.} \cite{Ammon:2010eq}.

In Appendix~\ref{app:InfiniteB}, we explore the well-known connection
between the light-front formalism and the system in the infinitely boosted frame. 
Susskind in Reference~\cite{Susskind:1967rg}
considered the system of free bosons in the infinitely boosted frame and
found that it is possible to consistently extract finite quantities which
found in non-relativistic systems.
In particular, the resulting Hamiltonian generates motion in one of the
light-like directions, {\it i.e.}, one of the light-cone coordinates is the
analog of the non-relativistic time.
Subsequently, Bardakci and Halpern \cite{Bardakci:1969dv} 
and Chang and Ma \cite{Chang:1968bh} argued that
the infinite-momentum limit is equivalent to adopting light-cone
coordinates.
Then we expect to extract the properties of Schr\"odinger
black holes from the counterparts of AdS black hole in the infinitely boosted frame.
This is demonstrated in the appendix for the case of the planar black
hole. By doing so we rederive the results of Section~\ref{sec:procedure}.

We exclusively work on the five dimensional AdS spacetime for concreteness.
However, the method adopted here to generate the quantities of
Schr\"odinger space can be trivially generalized to other dimensions.

%%%%%%%%%%%%%%%%%%%%%%%%%%%%%%%%%%%%%%%%%%%%%%%%%%%%%%%%%%%%%%%%%%%%%
\section{The Brown-York Procedure}\label{sec:procedure}
%%%%%%%%%%%%%%%%%%%%%%%%%%%%%%%%%%%%%%%%%%%%%%%%%%%%%%%%%%%%%%%%%%%%%
In this section, we carefully work out the Brown-York 
procedure \cite{Brown:1992br} for the planar AdS$_5$ black hole in
the special light-cone coordinates mentioned in the introduction.
This simple system serves as a prototype for more complex black holes.

The action of the gravitational system in concern is
\begin{align}\label{eq:OriginalAct}
  I = \frac{1}{16\pi G_5} \int d^5x\sqrt{-g}
  \bigg( \mathcal{R} + \frac{12}{R^2} \bigg)
  - \frac{1}{8\pi G_5}\int d^4x \sqrt{-\gamma}
  \bigg( K + \frac{3}{R} + \frac{R}{4}\mathcal{R}_4 \bigg)
  \;,
\end{align}
where the symbols $g$, $\mathcal{R}$ and $R$ are the determinant of the metric,
the scalar curvature and the length scale of the theory that is related to
the cosmological constant, respectively.
We have included the boundary terms and $\gamma$ denotes the determinant
of the boundary metric.
The first boundary term is the
Gibbons-Hawking term \cite{Gibbons:1976ue}
with the trace of the boundary second-fundamental form $K$, and the
second and third terms are the
standard five-dimensional counterterms of Balasubramanian and Kraus 
\cite{Balasubramanian:1999re} with the (intrinsic) scalar curvature
of the boundary $\mathcal{R}_4$.

The planar black hole solution to the equations of motion following
from the action can be written as
\begin{align}\label{eq:pOriginal}
  ds^2 = \bigg(\frac{r}{R}\bigg)^2 ( -h dt^2 + dx^2 + dy^2 + dz^2)
         + \bigg(\frac{R}{r}\bigg)^2 h^{-1} dr^2
  \quad\text{with}\quad
  h:=1-\frac{r_H^4}{r^4}
  \;,
\end{align}
where $r=r_H$ is the location of the horizon and we set the boundary
at some large (with respect to $r_H$) fixed value of $r$.
Now, we introduce the light-cone coordinates mentioned in the introduction,
\begin{align}\label{eq:bLC}
  x^+ =& b(t+x)
  \;,\quad
  x^- = \frac{1}{2b}(t-x)
  \;,
\end{align}
where we modified the scaling dimension of $b$ as $[b]=-1$ in the mass unit, 
and thus $[x^+] = -2$ and $[x^-] = 0$ to have manifest dynamical exponent z=2.  
The metric takes the form
\begin{align}\label{eq:planar}
  ds^2 =& \bigg(\frac{r}{R}\bigg)^2 \bigg\{ \frac{1-h}{4b^2}dx^{+2}
  -(1+h)dx^+dx^- + (1-h)b^2 dx^{-2} + dy^2 + dz^2 \bigg\}
  \nonumber\\
  &+ \bigg(\frac{R}{r}\bigg)^2 h^{-1} dr^2
  \;.
\end{align}
One can check that this is nothing but the boosted form of the metric
(\ref{eq:pOriginal}) transformed into usual light-cone coordinates 
with modified scaling dimensions of light-cone coordinates.
Thus, the parameter $b$ signifies the boosted system of the Lorentz
non-invariant solution, and in the end the metric (\ref{eq:planar}) describes non-relativistic 
setup with dynamical exponent z=2.

In the spirit of the infinite-boost limit of
Reference~\cite{Susskind:1967rg} mentioned in the introduction,
the non-relativistic ``limit'' corresponds to re-interpreting
one of the light-cone coordinates, say $x^+$, as the time.%
\footnote{
 In Appendix~\ref{app:InfiniteB}, we show how to derive the results
 from the infinite-boost limit, rather than adopting light-cone coordinates.
}
Then the ADM form of the metric is
\begin{align}\label{eq:planarADM}
  ds^2 =& \bigg(\frac{r}{R}\bigg)^2 \bigg\{ - \frac{h}{1-h}b^{-2} dx^{+2}
  + (1-h)b^2 \bigg( dx^- - \frac{1}{2b^2}\frac{1+h}{1-h}dx^+ \bigg)^2
  + dy^2 + dz^2 \bigg\}
  \nonumber\\
  &+ \bigg(\frac{R}{r}\bigg)^2 h^{-1} dr^2
  \;.
\end{align}
From the ADM form, we can read off the lapse function $N$, the shift function
$V^i$ and the horizon coordinate velocity in the $x^-$ direction
$\Omega_H$, which can be interpreted as chemical potential associated with the 
conserved quantities along the $x^-$ direction,  as
\begin{align}\label{eq:lapseShift}
  N = \bigg(\frac{r}{R}\bigg)\sqrt{\frac{h}{1-h}}\,b^{-1}
  \;,\quad
  V^- = - \frac{1}{2b^2}\frac{1+h}{1-h}
  \quad\text{and}\quad
  \Omega_H = \frac{1}{2b^2}
  \;.
\end{align}

Notice that we have alluded two kinds of hypersurfaces; the timelike
boundary surface at a large fixed $r$ and the spacelike
surface at a fixed time $x^+$ whose time development is described
by the ADM form.
Moreover, the spacelike surface at the intersection of those two
hypersurfaces plays a crucial role for the definition of
the Brown-York conserved quantities.
Since we have the non-trivial shift function, we must pay careful
attention to the projection tensors onto those surfaces.
The definition of the projections, of course, requires the normals
to the surfaces, and we take the unit normal of the timelike boundary
surface as
\begin{align}
  n_\mu := \bigg(\frac{R}{r}\bigg)h^{-1/2} (0,0,0,0,1)
  \;,
\end{align}
where the components are ordered according to $(+,-,y,z,r)$, and
following Brown and York \cite{Brown:1992br}, we define the normal
of the spacelike surface by
\begin{align}
  u_\mu := -N (1,0,0,0,0)
  \;,
\end{align}
where $N$ is the lapse function given in Equation~(\ref{eq:lapseShift}).
The projections onto the four-dimensional timelike boundary hyperspace
and the three-dimensional spacelike intersection surface are given
respectively as
\begin{align}
    \gamma_{\mu\nu} := g_{\mu\nu} - n_\mu n_\nu
  \quad\text{and}\quad
  \sigma_{\mu\nu} := g_{\mu\nu} - n_\mu n_\nu + u_\mu u_\nu
  \;.
\end{align}
Since they are projection operators, they do not have inverses
and the five-dimensional indices are raised and lowered by 
the metric $g_{\mu\nu}$.
However, if we restrict them to the appropriate components, namely,
$\gamma_{ij}$ with $i,j\in\{+,-,y,z\}$ and
$\sigma_{ab}$ with $a,b\in\{-,y,z\}$, 
then they have well-defined inverses and can be defined as
the metric on the respective surfaces.

The key object in deriving the Brown-York conserved quantities is
the stress-energy-momentum tensor.
It is defined as the on-shell value
of the variation
\begin{align}
  T_{ij} := - \frac{2}{\sqrt{-\gamma}}\frac{\delta I_{bd}}{\delta\gamma^{ij}}
  \;,
\end{align}
where $I_{bd}$ is the boundary terms in the action (\ref{eq:OriginalAct}).
Explicitly, we have
\begin{align}
  T_{ij} = \frac{1}{8\pi G_5} \bigg( K_{ij} - \gamma_{ij}K 
  - \frac{3}{R}\gamma_{ij} + \frac{R}{2}G_{4\,ij} \bigg)
  \;,
\end{align}
where $G_{4\,ij}$ is the Einstein tensor with respect to the metric
$\gamma_{ij}$.%
\footnote{
  The four-dimensional Einstein tensor obviously vanishes
  for our boundary of the planar black hole in discussion.
}
This $T_{ij}$ is evaluated at the boundary of the solution
(\ref{eq:planar}), and then the boundary is removed to infinity.
We remark that the projected second fundamental form $K_{ij}$ is defined as
\begin{align}\label{eq:2ndFF}
  K_{ij} := {\gamma_i}^\mu{\gamma_j}^\nu K_{\mu\nu}
  \quad\text{with}\quad
  K_{\mu\nu} := - n_{\nu;\mu}
  \;.
\end{align}

Given the stress-energy-momentum tensor, we can proceed to compute
the mass of the black hole as instructed by Brown and York.
We suggest that the appropriate mass for our system is not the quasilocal
energy, but the Hamiltonian that generates the unit time translation
\begin{align}
  M := \int d^3x \sqrt{\sigma} (N\epsilon-V^aj_a)
  \;,
\label{DefMass}  
\end{align}
where as defined in Reference~\cite{Brown:1992br}, we have
\begin{align}
  \epsilon := u_iu_jT^{ij}
  \quad\text{and}\quad
  j_a := - \sigma_{ai}u_jT^{ij}
  \;.
\label{DefDensity}  
\end{align}
The reader should pay careful attention to the indices where proper
projections must be done.
For example, $V^a$ is the projection of $V^i$ onto the intersection surface,
the indices of $T_{ij}$ are raised by the metric $\gamma_{ij}$ and also
we have $u_i:={\gamma_i}^\mu u_\mu$.
The momentum in the $x^-$ direction can be also computed following Brown
and York.
Taking the Killing vector field $\phi^a = (1,0,0)$, they tell us to compute
\begin{align}
  J = \int d^3x \sqrt{\sigma} j_a \phi^a
  \;,
\label{DefMomentum}  
\end{align}
which can be interpreted as the total particle number. 
We define the entropy, $S$, of the system as a quarter of the volume given at
the fixed time ($x^+$) and at the horizon ($r=r_H$).
The temperature can be obtained by requiring the smoothness of the Euclidean
geometry where the time $x^+$ is analytically continued to $ix^+$.
In doing so, the ADM form (\ref{eq:planarADM}) must be employed, as in the
case of the Kerr black hole.
Alternatively, the same result can be derived through the computation
of the surface gravity.

The thermodynamic quantities computed as described are identical to
the ones derived from the planar Schr\"odinger black hole
\cite{Herzog:2008wg,Maldacena:2008wh,Yamada:2008if},
namely,
\begin{align}\label{eq:pThermo}
  \Omega_H = \frac{1}{2b^2}
  \;,\quad
  M = \frac{r_H^4V_3}{16\pi G_5R^5}
  \;,\quad
  J = - \frac{r_H^4V_3}{4\pi G_5R^5}b^2
  \;,\quad
  S = \frac{r_H^3V_3}{4G_5R^3} b
  \;,\quad
  \beta = \frac{\pi R^2}{r_H} b
  \;,
\end{align}
where we have defined $V_3 := \int dx^-dydz$.
However, we would like to emphasize that we have computed the quantities
through the usual well-defined procedure, as the IAS group has done so
already in Reference~\cite{Maldacena:2008wh}. 
On the other hand, the other derivations involve
ambiguous counterterms and necessity to assume the validity of 
the first law, rather than independently checking it.
Also notice the simplicity of the method, as compared to the ones that
are employed for the Schr\"odinger black holes.
Taking advantage of the simplicity, we are going to apply this method
to the complicated systems of R-charged black holes in the next section.

%%%%%%%%%%%%%%%%%%%%%%%%%%%%%%%%%%%%%%%%%%%%%%%%%%%%%%%%%%%%%%%%%%%%%
\section{Thermodynamics of R-Charged
Black Holes}\label{sec:RCharged}
%%%%%%%%%%%%%%%%%%%%%%%%%%%%%%%%%%%%%%%%%%%%%%%%%%%%%%%%%%%%%%%%%%%%%
This section can be safely skipped for the reader who is more interested
in the computation of the conductivity in Section~\ref{sec:DC}. 

The R-charged black holes have three independent charges in general. 
The Schr\"odinger version with the three equal charges
has been obtained in
References~\cite{Imeroni:2009cs,Adams:2009dm} through the complicated
null Melvin twist.
The equal-charge configuration is a very special case because the scalars
decouple from the theory and the effective five dimensional action
becomes simple Einstein-Maxwell with the negative cosmological constant
\cite{Chamblin:1999tk}.
Applying the null Melvin twist to the general-charge configuration is a
very cumbersome task.
Here, we see that the thermodynamic properties
can be obtained easily for the general-charge configuration
by adopting the light-cone coordinates
in the AdS R-charged black hole systems.
There are planar, spherical and hyperbolic black hole solutions in the AdS space
and we are going to introduce them simultaneously.
Then we discuss the phase diagrams of the planar and spherical black
holes.

\bigskip

One way to describe the five-dimensional R-charged black holes
is to regard them as the solutions of the five dimensional $\mathcal{N}=2$
gauged $U(1)^3$ supergravity \cite{Behrndt:1998jd}. 
The bosonic part of the action is
\begin{align}\label{eq:RAction}
  I = &\frac{1}{16\pi G_5} \int d^5x\sqrt{-g}
        \bigg[ \mathcal{R} 
        - \frac{1}{2}\sum_{i=1}^3
               (\partial_{\mu} \ln X_i)(\partial^{\mu} \ln X_i)
        - \frac{1}{4}\sum_{i=1}^3 X_i^{-2}
                          F_{i\,\mu\nu}{F_i}^{\mu\nu}
        + \frac{V}{R^2} \bigg]
        \nonumber \\
        &+\frac{1}{16\pi G_5}
         \int F_1\wedge F_2\wedge A_3
  - \frac{1}{8\pi G_5}\int d^4x \sqrt{-\gamma}
  \bigg( K + \frac{3}{R} + \frac{R}{4}\mathcal{R}_4
  + \frac{1}{2R}\vec\phi^2 \bigg)
  \;.
\end{align}
The fields $X_{i}$ with $i=1,2,3$ are the scalars
and they are subject to the constraint
\begin{align}
  X_1X_2X_3 \equiv 1
        \;,
\end{align}
so there are actually two independent scalar fields.
The potential $V$ is defined as
\begin{equation}
  V := 4 \sum_{i=1}^3 X_i^{-1}
        \;.
\end{equation}
The one-form fields $A_i$ correspond to
the $U(1)^3$ gauged symmetry and $F_i:=dA_i$.
As in Section~\ref{sec:procedure}, we have included the boundary terms
but with extra $\vec\phi^2$, which is the Liu-Sabra finite counterterm
\cite{Liu:2004it} and this will be defined shortly.

The black hole solutions to the equations of motion that follow from the action
can be written as
\begin{align}
  ds^2 =& \bigg(\frac{r}{R}\bigg)^2 H(r)^{1/3}
  \big\{ -h(r)dt^2 + \eta_k^2 + dX_k^2 \big\}
  + \bigg(\frac{R}{r}\bigg)^2 H(r)^{-2/3} h(r)^{-1} dr^2
  \;,\nonumber\\
  X_i(r) =& H(r)^{1/3}/H_i(r)
  \quad\text{and}\quad
  A_i(r) = \bigg( \frac{g_i}{r_H^2+q_i} - \frac{g_i}{r^2+q_i} \bigg)dt
  \;.
\end{align}
The function $H$ is
\begin{align}
  H = H_1 H_2 H_3
  \quad\text{with}\quad
  H_i := 1 + \frac{q_i}{r^2}
  \;,
\end{align}
where $q_i$ are the parameters related to the $U(1)^3$ charges $g_i$ via
\begin{align}
  g_i = \sqrt{q_i (r_0^2+kq_i)}
  \;.
\end{align}
The parameter $r_0$ is the non-extremality parameter and the integer $k$
will be explained momentarily.
The blackening factor $h(r)$ is given by
\begin{align}
  h := 1+ \bigg(\frac{R}{r}\bigg)^2
  \bigg\{k-\bigg(\frac{r_0}{r}\bigg)^2\bigg\} H^{-1}
  \quad\text{and}\quad
  r_0 = r_H \sqrt{k+\bigg(\frac{r_H}{R}\bigg)^2H_H}
  \;,
\end{align}
where the parameter $r_H$ is the location of the horizon in the $r$ coordinate
and is the largest root of $h=0$.
We have expressed the non-extremality parameter in terms of $r_H$ and
the short hand  notation $H_H:=H(r_H)$ is introduced.
The parameter $k$ represents three possible values, $k=0$, $+1$ and $-1$, and
they correspond to the planar, spherical and hyperbolic black holes,
respectively.
According to the value of the parameter $k$, we have
\begin{align}
  \eta_0 :=& dx
  \quad\text{and}\quad
  dX_0^2 := dy^2 + dz^2
  \;,\nonumber\\
  \eta_{+1} :=& \frac{R}{2} \big( d\psi + \cos\theta d\phi \big)
  \quad\text{and}\quad
  dX_{+1}^2 := \bigg(\frac{R}{2}\bigg)^2 \big( d\theta^2
  + \sin^2\theta d\phi^2 \big)
  \;,\nonumber\\
  \eta_{-1} :=& \frac{R}{2} \big( d\psi + \cosh\chi d\phi \big)
  \quad\text{and}\quad
  dX_{-1}^2 := \bigg(\frac{R}{2}\bigg)^2 \big( d\theta^2
  - \sinh^2\chi d\phi^2 \big)
  \;.
\end{align}
The one-forms $\eta_{\pm 1}$ are introduced in Reference~\cite{Yamada:2008if}
for the Schr\"odinger black holes.

Following that reference, we introduce the coframe
\begin{align}
  \omega^+ := b(dt+\eta_k)
  \;,\quad
  \omega^- := \frac{1}{2b}(dt-\eta_k)
  \;,\quad
  \vec\omega_k
  \quad\text{and}\quad
  \omega^r = dr
  \;,
\end{align}
where
\begin{align}
  \vec\omega_0 := (dy,dz)
  \;,\quad
  \vec\omega_{+1} := (\frac{R}{2}d\theta, \frac{R}{2}\sin\theta d\phi)
  \quad\text{and}\quad
  \vec\omega_{-1} := (\frac{R}{2}d\chi, \frac{R}{2}\sinh\chi d\phi)
  \;.
\end{align}
For the cases $k=\pm 1$, they are non-coordinate bases and their
dual frames, $\{e_\mu\}$,
and Lie brackets are given in Appendix~\ref{app:nonCoords}.
Also recall that the normalization of $\omega^\pm$ with the parameter
$b$ is the boost for the planar case.
This physical interpretation clearly fails when $k=\pm 1$, because
$\eta_{\pm 1}$ are not translationally invariant directions.
We comment them further when we discuss their thermodynamics separately.

In these bases, the metric takes the form
\begin{align}\label{eq:genRChargedMet}
  ds^2 =& \bigg(\frac{r}{R}\bigg)^2 H^{1/3} 
  \bigg\{ \frac{1-h}{4b^2}\omega^{+2}
  -(1+h)\omega^+\omega^- + (1-h)b^2 \omega^{-2} + \vec\omega_k^2 \bigg\}
  \nonumber\\
  &+ \bigg(\frac{R}{r}\bigg)^2 H^{-2/3} h^{-1} \omega^{r2}
  \;,
\end{align}
where $\vec\omega_k^2$ are defined as
\begin{align}
  \vec\omega_0^2 := \omega^{y2}+\omega^{z2}
  \;,\quad
  \vec\omega_{+1}^2 := \omega^{\theta2}+\omega^{\phi2}
  \quad\text{and}\quad
  \vec\omega_{-1}^2 := \omega^{\chi2}-\omega^{\phi2}
  \;.
\end{align}
One notices that the coframes are chosen so that the metric
appears similar to the one in Equation~(\ref{eq:planar}).
As in Section~\ref{sec:procedure} and in Reference~\cite{Yamada:2008if},
we propose that the non-relativistic counterpart is obtained by
re-interpreting $\omega^+$ as the non-relativistic time direction.
Then the ADM form is
\begin{align}\label{eq:ADMgeneralRCharge}
  ds^2 =& \bigg(\frac{r}{R}\bigg)^2 H^{1/3}
  \bigg\{ - \frac{h}{1-h}b^{-2} \omega^{+2}
  + (1-h)b^2 \bigg( \omega^- - \frac{1}{2b^2}\frac{1+h}{1-h}\omega^+ \bigg)^2
  + \vec\omega_k^2 \bigg\}
  \nonumber\\
  &+ \bigg(\frac{R}{r}\bigg)^2 H^{-2/3} h^{-1} \omega^{r2}
  \;.
\end{align}
From this, we obtain the lapse function, the shift function and
the horizon coordinate velocity in the $\omega^-$ direction
\begin{align}\label{eq:RChargedLapseShift}
  N = \bigg(\frac{r}{R}\bigg)H^{1/6}\sqrt{\frac{h}{1-h}}\,b^{-1}
  \;,\quad
  V^- = - \frac{1}{2b^2}\frac{1+h}{1-h}
  \quad\text{and}\quad
  \Omega_H = \frac{1}{2b^2}
  \;.
\end{align}

We are almost ready to compute the thermodynamic quantities, except
that the Liu-Sabra finite counterterm in the action (\ref{eq:RAction})
needs to be defined.
This is the term introduced in Reference~\cite{Liu:2004it}, and unlike other
counterterms, it is not necessary to cancel the divergences of
the on-shell action.
Being plainly square of the two (independent) scalar fields
$\vec\phi=(\phi_1,\phi_2)$,
this is the simplest finite non-vanishing
matter field term. It can be explicitly written in terms of
the functions $H_i$ as
\begin{align}
  \vec\phi^2 = \frac{1}{6} \bigg\{ 3 \bigg(\ln\frac{H_1}{H_2}\bigg)^2
  + \bigg(\ln\frac{H_1H_2}{H_3^2} \bigg)^2 \bigg\}
  \;.
\end{align}
It is commonly considered that the addition of finite counterterms corresponds
to changes in the renormalization scheme in the dual field theory.
However, our main motivation to include this term comes from the fact that
only {\it with the} finite counterterm, does the Brown-York procedure yield
thermodynamic quantities that satisfy the first law.

%%%%%%%%%%%%%%%%%%%%%%%%%%%%%%%%%%%%%%%%%%%%%%%%%%%%%%%%%%%%%%%%%%%%%
\subsection{Thermodynamic Quantities}
The thermodynamic quantities of the R-charged black holes are obtained in
a completely similar manner as in Section~\ref{sec:procedure}.
The masses and momenta are calculated by the Brown-York procedure
with the normal vectors of the hypersurfaces and the projection tensors
defined similarly as before.
For the cases $k=\pm 1$, the stress-energy-momentum tensors should be
computed with the geometric quantities in the non-coordinate bases 
summarized in Appendix~\ref{app:nonCoords}.
The entropy and temperature are computed from the metric expressed
in the special coframes above.

In addition, we have the $U(1)^3$ R-charges and the conjugate chemical
potentials.
The charges $N_i$ are computed through Gauss' law,
\begin{align}
  N_i = \lim_{r\to\infty}-\frac{1}{16\pi G_5}\int \omega^{\wedge 3}\sqrt{\sigma}
  n_\mu u_\nu F_i^{\mu\nu}
  \;,
\end{align}
where $n_\mu$ and $u_\nu$ are the normals of the timelike and spacelike
hypersurfaces introduced in Section~\ref{sec:procedure}, and
$\omega^{\wedge 3}:=\omega^-\wedge\omega^y\wedge\omega^z$ for the $k=0$ case
and similarly defined for the $k=\pm 1$ cases.
Recall that Gauss' law follows from the action by taking the variation
of the Lagrange multiplier, {\it i.e.}, the time component of the gauge
fields, $A_ie_+$.
This immediately implies that the conjugate chemical potentials, $\mu_i$,
are exactly the multipliers.
However, there are two issues to which we must pay attention.
First, the chemical potentials must be the {\it difference} of
the time component $A_ie_+$ between the boundary and the horizon.
Since we have set $A_i(r_H)=0$ as required by the regularity of the vector
field, the chemical potential should be the value at $r=\infty$.
The second issue is that our system has the horizon velocity
$\Omega_H=1/(2b^2)$, so it is more adequate to write the gauge fields as
\begin{align}\label{eq:gaugeField}
  A_i(r=\infty) = \frac{g_i}{r_H^2+q_i}
  \big\{ b^{-1}\omega^+ + b( \omega^- - \frac{1}{2b^2}\omega^+ ) \big\}
  \;,
\end{align}
and take the first term as the chemical potentials.
This definition of the chemical potentials is equivalent to the (difference)
value of the gauge fields in the co-moving frame with respect to the horizon.
Changing to the co-moving frame is necessary to avoid 
the ill behavior of the timelike killing vector
field and the associated time component of
the gauge fields inside the ergo-region,
whose existence can be clearly seen in the ADM form
(\ref{eq:ADMgeneralRCharge}).%
\footnote{
  More discussion on this point can be found in
  Reference~\cite{Yamada:2008em}.
  As will be mentioned shortly, this issue is overlooked
  in Reference~\cite{Imeroni:2009cs} and consequently, there is an
  extra factor of $2$ in their chemical potential.
  In Reference~\cite{Adams:2009dm}, this chemical potential is
  not discussed.
}

We collect the thermodynamic quantities described above;
\begin{align}
  \Omega_H =& \frac{1}{2b^2}
  \;,\quad
  M = \frac{V_3}{16\pi G_5R^3}
  \bigg( r_0^2 + \frac{2}{3}k\sum_iq_i + |k|\frac{R^2}{4} \bigg)
  \nonumber\\
  J =& -\frac{V_3}{4\pi G_5R^3}
  \bigg( r_0^2 + \frac{2}{3}k\sum_iq_i + |k|\frac{R^2}{4} \bigg) b^2
  \;,\quad
  S = \frac{r_H^3H_H^{1/2}V_3}{4G_5 R^3} b
  \nonumber\\
  \beta =& \frac{H_H^{1/2}}{Q_k}\frac{\pi R^2}{r_H} b
  \;,\quad
  N_i = \frac{g_iV_3}{8\pi G_5R^3} b
  \;,\quad
  \mu_i = \frac{g_i}{r_H^2+q_i} b^{-1}
  \;,
\end{align}
where we have defined $V_3:=\int\omega^{\wedge 3}$ and
\begin{align}
  Q_k := 1 + \frac{kR^2 +q_1+q_2+q_3}{2r_H^2}
  - \frac{q_1q_2q_3}{2r_H^6}
  \;.
\end{align}
One can check that the quantity
\begin{align}\label{eq:ThermAction}
  \beta\bigg(M - \beta^{-1}S - \Omega_HJ - \sum_i\mu_iN_i \bigg)
  = \frac{(\beta V_3)\,r_H^2}{16\pi G_5R^3}
  \bigg\{ k + \frac{3}{4}|k|\bigg(\frac{R}{r_H}\bigg)^2
  - \bigg(\frac{r_H}{R}\bigg)^2H_H \bigg\}
  \;,
\end{align}
precisely is the on-shell value of the action with the boundary removed to
infinity, and this serves as a non-trivial check on our procedure.

\bigskip

As noted in the foregoing sections, 
our derivation of the thermodynamic quantities
do not assume the first law, while the previous derivations do depend
on it and the independent check of the law has not been possible.
Therefore it is a worthwhile digression here to discuss our results
in comparison with the known ones.

The planar case without the charges was discussed already
in Section~\ref{sec:procedure}, 
and our results are identical to the previously derived quantities.
Imeroni and Sinha in Reference~\cite{Imeroni:2009cs} discuss a special
charged case with $(q_1,q_2,q_3)=(q,q,q)$.%
\footnote{
  To compare their results with ours, one must bring their charge $Q$ to
  our $g=g_i$ (for all $i$) and their $r$ to our $\sqrt{r^2+q}$ (consequently
  their $r_0$ to $\sqrt{r_H^2+q}$).
}
Their derivation involves type IIB supergravity action with unconventional
counterterms.
The counterterms are not unique and
the coefficients are determined through a number of (reasonable) assumptions.
The thermodynamic quantities are obtained from the regulated action, by
effectively assuming the first law.
However, they identify the entropy as a quarter of the horizon area
with the original relativistic coordinates ($t$ and $x$), 
and this is not an appropriate identification for 
the non-relativistic
system with the time direction $\omega^+$.%
\footnote{
  They identify the temperature, the horizon velocity and
  the chemical potential
  using the coordinate system $\omega^\pm$, so it is inconsistent.
}
Moreover, they define the chemical potential without taking into account of
the horizon velocity and the existence of the ergo-region, which is
not adequate for the system.
These lead to the mismatch with our results, but
their results agree with ours after correcting the entropy and the chemical
potential.

The spherical and hyperbolic cases without the charges were discussed in
Reference~\cite{Yamada:2008if}.
The derivation of the on-shell action was based on the background subtraction
method with an unusual boundary matching, which is highly {\it ad hoc}.
The thermodynamic quantities were derived, again, by effectively assuming
the first law.
However, in our thermodynamic quantities with $k=\pm 1$, the first law is
{\it not} satisfied with respect to the variable $b$.
From the viewpoint of our procedure, this is reasonable, because
the interpretation of the variable $b$ as a boost parameter is
only possible for the $k=0$ case.
Hence the parameter $b$ (for $k=\pm 1$)
is not the variable that determines the thermodynamic
equilibrium.
In other words, we cannot take the variation with respect to $b$
to extremize the action.
This fact leads to the discrepancy between our results and those of
Reference~\cite{Yamada:2008if}, because the latter {\it assumes}
that $b$ is a variable for the extremization.
We are unable to determine which is the correct identification of the
thermodynamic quantities, for the methods are completely different.
However, we will see that the emerging phase diagrams turn out to be
identical (for the uncharged setting).

%%%%%%%%%%%%%%%%%%%%%%%%%%%%%%%%%%%%%%%%%%%%%%%%%%%%%%%%%%%%%%%%%%%%%
\subsection{Planar Black Hole Phase Diagram}\label{sec:k0phase}
Let us specialize to the planar black hole (with $k=0$) and discuss
its phase structure.
As usual in the system of AdS black holes, the first interest is
the phase transition between the black hole and the AdS space without
the black hole, {\it i.e.}, the generalization of the Hawking-Page
phase transition \cite{Hawking:1982dh} to the finite charges and
chemical potentials.
The phases are determined by comparing the values of the action
evaluated on the respective solutions.
The on-shell value of the action for the black hole is given in
Equation~(\ref{eq:ThermAction}).
The other solution of interest can be obtained by setting
$r_0=0=q_i$ in the black hole solution, and
the value of the action (\ref{eq:RAction}) with respect to
this solution turns out to be zero.
Since the black hole solution always gives negative on-shell action,
the black hole is always preferred and there is no Hawking-Page
phase transition for this planar case.

\begin{figure}[ht]
{
\centerline{\scalebox{1.2}{\includegraphics{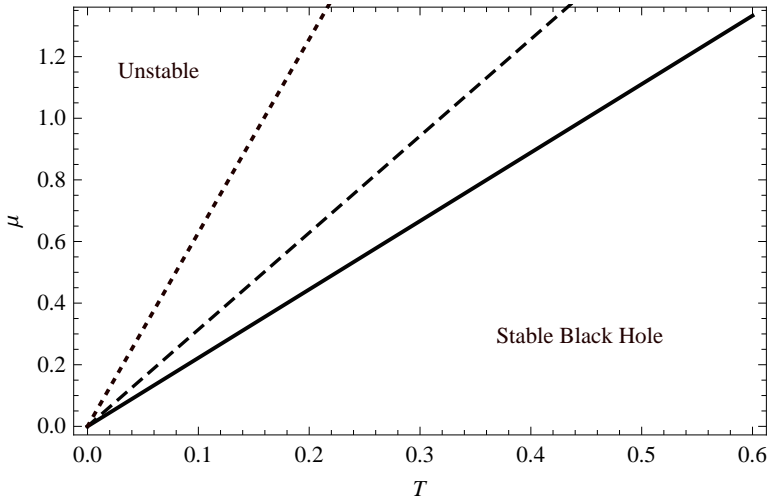}}}
\caption{\footnotesize
  The stability threshold lines in $T$-$\mu$ phase diagram.
  The solid, dashed and dotted lines are for the configurations
  $(q,0,0)$, $(q,q,0)$ and $(q,q,q)$, respectively.
  The slops are $\pi/\sqrt{2}$, $\pi$ and $2\pi$.
  We have set $R=1$.
}\label{fig:k0Phase}%
}
\end{figure}

Another interesting structure in the phase diagram is the thermodynamic
stability threshold.
We determine the threshold line by following Reference~\cite{Cvetic:1999ne}.
We calculate the Hessian matrix of the left-hand side in
Equation~(\ref{eq:ThermAction}) with respect to the variables
$r_H$, $q_i$ and $b$, but with fixed $\beta$, $\Omega_H$ and $\mu_i$.
Then the determinant of the matrix is evaluated with the on-shell
values of the fixed quantities, and the zero of this quantity
determines the threshold.
We have carried out this analysis for the charge configurations
$(q,0,0)$, $(q,q,0)$ and $(q,q,q)$.%
\footnote{
  For the case $(q,q,0)$, it is important to note that the Hessian
  must be computed with respect to the two charge parameters $q_1$ and
  $q_2$, and then compute the determinant of it with the constraint
  $q:=q_{1,2}$.
  Similarly for the configuration $(q,q,q)$.
}
The analysis reveals that the stability thresholds in the $T$-$\mu$
phase diagram are given as straight lines, as shown in Figure~\ref{fig:k0Phase}.
The phase diagram is identical to the relativistic case, as worked
out in Reference~\cite{Yamada:2008em}.
It turns out that the critical lines are independent of
the parameter $b$, and as long as $\mu$ and $T$ are in the stable
region, any value of $b$ is allowed.
The similarity of the Schr\"odinger black hole properties to the
relativistic counterpart has been
pointed out since References~\cite{Herzog:2008wg,Maldacena:2008wh},
and the situation does not change with the inclusion of the R charges.

%%%%%%%%%%%%%%%%%%%%%%%%%%%%%%%%%%%%%%%%%%%%%%%%%%%%%%%%%%%%%%%%%%%%%
\subsection{Spherical Black Hole Phase Diagram}\label{sec:kp1phase}
Let us now focus on the spherical black hole with $k=+1$.
We can determine the Hawking-Page phase transition as described in
the previous subsection. We have the difference action
\begin{align}
  \Delta I = \frac{(\beta V_3)r_H^2}{16\pi G_5R^3}
  \bigg( 1 - \frac{r_H^2}{R^2}H_H \bigg)
  \;.
\end{align}
When $\Delta I$ is negative, the black hole is preferred to
the AdS space without the black hole, and when it is positive,
the preference is the other way around.
The thermodynamic stability can also be analyzed as before using 
the Hessian matrix with respect to the variables
$r_H$ and $q_i$.

\begin{figure}[ht]
{
\centerline{\scalebox{1.2}{\includegraphics{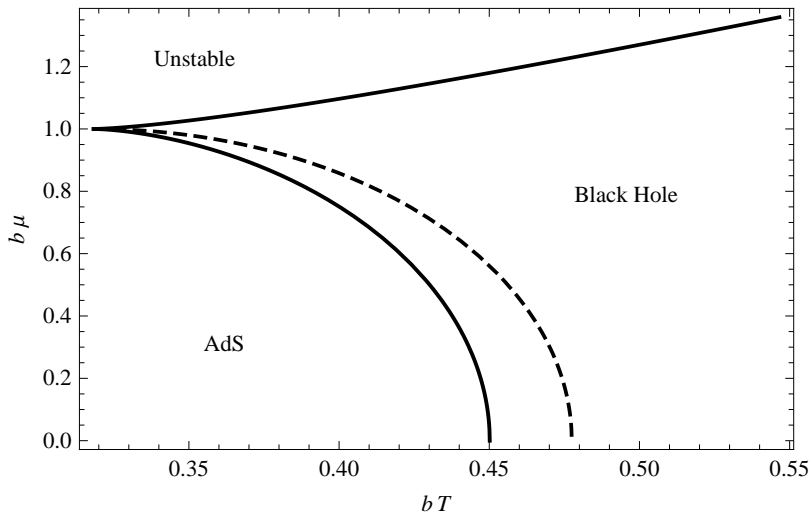}}}
\caption{\footnotesize
  The $k=+1$ phase diagram in $T$-$\mu$ parameter space
  with the charge configuration $(q,0,0)$.
  The temperature and the chemical potential are scaled
  with the parameter $b$.
  The solid lines are the stability thresholds and the dashed
  curve is the Hawking-Page phase transition line.
  The curves merge and terminate at $(bT,b\mu)=(1/\pi,1)$.
  We have set $R=1$.
}\label{fig:k1Tmuq}%
}
\end{figure}

In Figure~\ref{fig:k1Tmuq}, the $T$-$\mu$ phase diagram for
the charge configuration $(q,0,0)$ is shown.
This is identical to the relativistic counterpart as worked
out in Reference~\cite{Cvetic:1999ne}.(See also \cite{Yamada:2008em}.)

The curves in the diagram merge and terminate at $(bT,b\mu)=(1/\pi,1)$
in the units of $R$.
From the discussion so far, the region of the phase diagram
$bT<1/\pi$ is completely unclear because the black hole saddle point
in the action does not exist.
However, it is shown in Reference~\cite{Yamada:2008em} that there is
a metastability line running at $b\mu=1$ for all values of $T$, and
it is argued that the metastability line below $bT<1/\pi$ is
the natural continuation of the stability threshold, but representing
the stability of the AdS space without the black hole.
The physical cause of
this instability is hard to see in the five dimensional action
(\ref{eq:RAction}).
However, when describing the action as the $S^5$ compactification
of type IIB supergravity with rotations in the five sphere,
the critical line corresponds to the point where the speed of
the rotation exceeds the speed of light \cite{Chamblin:1999tk,Cvetic:1999ne}.
Thus it appears that the thermodynamics
of the five dimensional theory reflects the misbehavior in the
higher dimensional theory.
The instability line at $b\mu=1$
can also be demonstrated from the dual field theory side.
The field theory dual of the critical line at $b\mu=1$ is where the chemical
potentials become greater than the mass of the scalar fields, induced by
the conformal coupling of the fields to the curvature of the space, $S^3$.
This phenomenon is shown in Reference~\cite{Yamada:2006rx}
for the weak coupling
case and the authors argue that the critical line persists to the strong
coupling regime due to the supersymmetry.
Therefore, in what follows, we assume the critical line at $b\mu=1$
for all values of temperature.

The $T$-$\mu$ phase diagram of the charge configuration $(q,q,0)$ is
similar to the one-charge configuration just discussed, except that
the curves of the phase diagram merge and terminate at
$(bT,b\mu)=(1/2\pi,1)$.
The configuration $(q,q,q)$ also have a similar phase diagram but as
is well-known, the black hole with this charge configuration can become
zero temperature with finite entropy, so the curves merge
at $(bT,b\mu)=(0,1)$.

\begin{figure}[ht]
{
\centerline{\scalebox{1.2}{\includegraphics{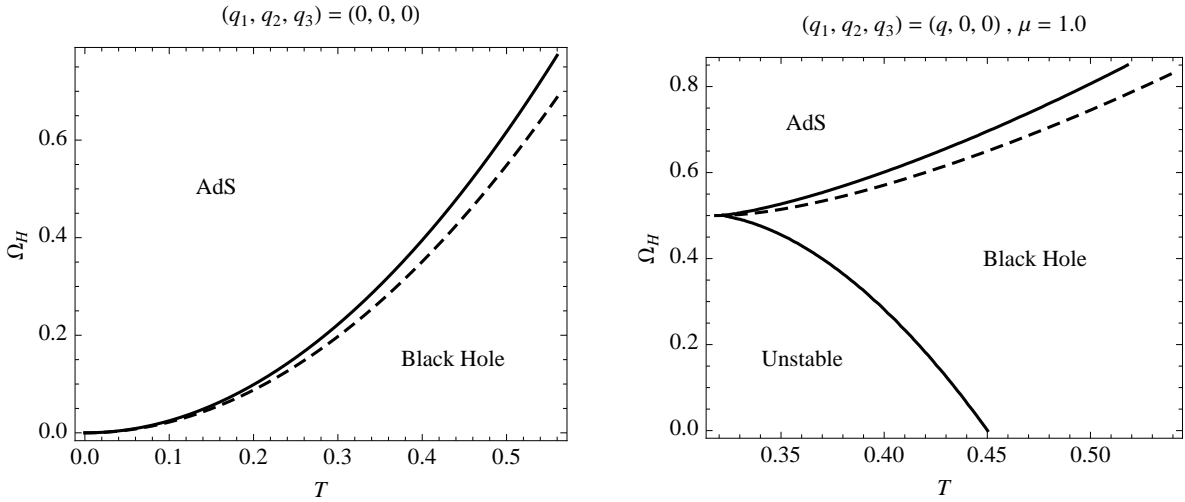}}}
\caption{\footnotesize
  The $k=+1$ phase diagram in $T$-$\Omega_H$ parameter space.
  The left diagram is the uncharged case and the right is for the
  charge configuration $(q,0,0)$ with $\mu=1$, where we have set $R=1$.
  The solid lines are the stability thresholds and the dashed
  curve is the Hawking-Page phase transition line.
  The curves merge at $(T,\Omega_H)=(0,0)$ for the left diagram and
  at $(1/\pi,0.5)$ for the right.
}\label{fig:k1TOm}%
}
\end{figure}

Let us now consider the $T$-$\Omega_H$ phase diagram with fixed $\mu$.
In Figure~\ref{fig:k1TOm}, the phase diagrams for the uncharged case
and the charged case with $(q_1,q_2,q_3)=(q,0,0)$ are shown.
For the uncharged black hole phase diagram,
the Hawking-Page phase transition and
the stability threshold are given by the curves 
$\Omega_H = (2\pi^2/9)T^2$ and $\Omega_H = (\pi^2/4)T^2$,
respectively.
(We are adopting the units of $R$.)
As mentioned before,
this phase diagram is identical to the one in Reference~\cite{Yamada:2008if},
despite the differences in each thermodynamic quantities.

For the charged case,
recall that we have the threshold at $bT=1/\pi$ and $b\mu=1$,
that is, at $T=\sqrt{2\Omega_H}/\pi$ and $\mu=\sqrt{2\Omega_H}$.
In the right diagram of Figure~\ref{fig:k1TOm}, we have chosen $\mu=1.0$
as an example, then the critical value of $\Omega_H$ is $0.5$ and
the lowest possible temperature is $1/\pi$, as one sees in
the diagram.
Similar phase diagrams are observed for the other charge
configurations $(q,q,0)$ and $(q,q,q)$.

\begin{figure}[ht]
{
\centerline{\scalebox{1.3}{\includegraphics{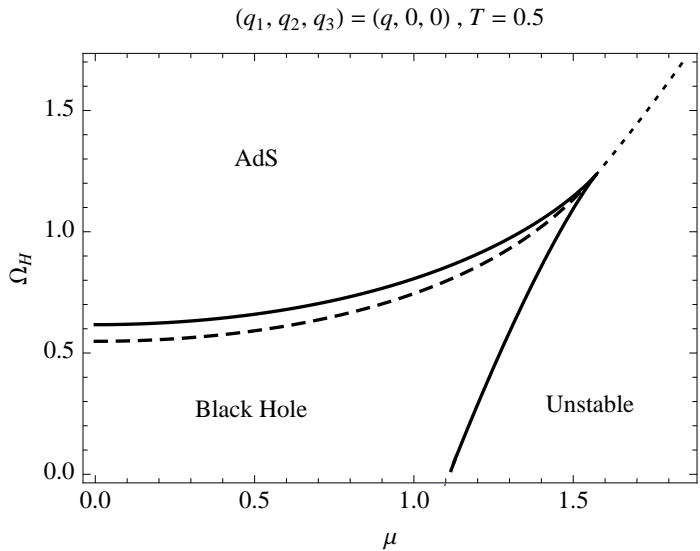}}}
\caption{\footnotesize
  The $k=+1$ phase diagram in $\mu$-$\Omega_H$ parameter space.
  The charge configuration is $(q,0,0)$ and we chose $T=0.5$, 
  where we have set $R=1$.
  The curves merge at $(\mu,\Omega_H)=(\pi/2,\pi^2/8)$ for the temperature
  chosen.
  The solid lines are the stability thresholds, the dashed
  curve is the Hawking-Page phase transition line and the dotted curve
  is the critical line $b\mu=1$, {\it i.e.}, $\Omega_H=\mu^2/2$.
}\label{fig:k1muOm}%
}
\end{figure}

We now consider the $\mu$-$\Omega_H$ phase diagram with fixed $T$.
Figure~\ref{fig:k1muOm} shows the phase diagram for the charge
configuration $(q,0,0)$.
For this plot, we have chosen $T=0.5$.
Then, translating the critical point $(bT,b\mu)=(1/\pi,1)$, we deduce
that the curves merge at $(\mu,\Omega_H)=(\pi/2,\pi^2/8)$, as observed
in the diagram.
We have also plotted the critical line $b\mu=1$, {\it i.e.},
$\Omega_H=\mu^2/2$ after the other curves merge and terminate.
The case $(q,q,0)$ is similar to Figure~\ref{fig:k1muOm} and
for $(q,q,q)$, the curves merge at infinity along the critical
curve $\Omega_H=\mu^2/2$.

%%%%%%%%%%%%%%%%%%%%%%%%%%%%%%%%%%%%%%%%%%%%%%%%%%%%%%%%%%%%%%%%%%%%%
\section{DC Conductivity}\label{sec:DC}
%%%%%%%%%%%%%%%%%%%%%%%%%%%%%%%%%%%%%%%%%%%%%%%%%%%%%%%%%%%%%%%%%%%%%

In this section, we demonstrate that the introduction of the light-cone 
coordinates (\ref{eq:bLC}) is enough to compute non-relativistic
quantities other than the thermodynamic ones presented in the previous sections.
In particular, we compute a two point correlator and the DC conductivity.

Since the planar black hole horizon has translationally invariant directions,
it is natural to consider two-point correlators and shear viscosity 
in terms of their Fourier modes,
following the prescription of Son and Starinets \cite{Son:2002sd}.
The computation is similar to theirs and sketched in 
Appendix~\ref{app:ratio} with the same results as the ones 
previously derived
from the Schr\"odinger black hole 
\cite{Herzog:2008wg,Maldacena:2008wh,Adams:2008wt}.
In particular, the viscosity-entropy ratio is identical to
the relativistic counterpart.

As another demonstration of the utility we consider the DC
conductivity, which is an interesting physical observables in 
condensed matter system such as high $T_c$ superconductor. At optimal doping, 
resistivity reveals an intriguing universal behavior as $\rho \sim T$. 
In holographic approach, there exist a few systems which show this 
interesting property. These include the AdS \cite{Faulkner:2010zz}, 
Lifshitz \cite{Hartnoll:2009ns} and charged dilatonic black holes 
\cite{Charmousis:2010zz}.  

In this section we work out the DC conductivity,
which is the non-relativistic counterpart of the computation by
Karch and O'Bannon~\cite{Karch:2007pd}.
The non-relativistic DC conductivity from the Schr\"odinger
black hole spacetime has been obtained in
Reference~\cite{Ammon:2010eq} and we shall compare the results.
For the sake of the comparison, the discussion here closely
follows Reference~\cite{Ammon:2010eq}.

We start with the full (asymptotically) AdS$_5\times S^5$ spacetime
expressed in the light-cone coordinates (\ref{eq:bLC})
\begin{align}
  ds^2 &= g_{++} dx^{+2} + 2 g_{+-} dx^+ dx^- + g_{--} dx^{-2}  
  + g_{yy} d y^2 + g_{zz} dz^2 \nonumber \\ 
  &+ g_{rr} dr^2 + \left(d\chi + \mathcal{A} \right)^2 + ds^2_{\cp} \;, 
  \label{eq:fullads5s5metricinconductivity}
\end{align}
where the AdS part of the metric is given in (\ref{eq:planar})
and the $S^5$ metric is expressed as 
a Hopf fibration over $\cp$, with $\chi$ the Hopf fiber direction.
The one-form
$\mathcal{A}$ gives the K\"ahler form $J$ of $\cp$ via $d\mathcal{A}=2J$. 
To write the metric of $\cp$ and $\mathcal{A}$ explicitly, we introduce $\cp$ 
coordinates $\alpha_1$, $\alpha_2$, $\alpha_3$, and $\theta$ and define 
the $SU(2)$ left-invariant forms
\begin{align}
	&\sigma_1 :=   \frac{1}{2} \left( \cos \alpha_2 \, d\alpha_1 
	+ \sin\alpha_1 \, \sin\alpha_2 \, d\alpha_3 \right) \;, 
	\nonumber \\ 
	&\sigma_2 := \frac{1}{2} \left( \sin \alpha_2 \, 
	d\alpha_1 - \sin\alpha_1 \, \cos\alpha_2 \, d\alpha_3 \right) \;, 
	\nonumber \\ 
	&\sigma_3 := \frac{1}{2} \left( d\alpha_2 + \cos\alpha_1 \, d\alpha_3\right) \;,
\end{align}
so that the coordinate expression of $\cp$ and $\mathcal{A}$ are
\begin{align}
	ds^2_{\cp} = d\theta^2 + \cos^2\theta \left( \sigma_1^2 
	+ \sigma_2^2 + \sin^2\theta \, \sigma_3^2 \right) \;,
	\quad  \mathcal{A} = \cos^2 \theta \, \sigma_3 \ .
\end{align}

As in References~\cite{Karch:2007pd,Ammon:2010eq}, we introduce $N_f$
probe D7-branes --- but in the AdS$_5\times S^5$ expressed in light-cone coordinates.
We are working with the probe limit, $N_f \ll N_c$, which suppresses the back-reaction 
of the D7-branes on the gravity background.

The D7-branes are extended to the AdS$_5$ and the angular directions 
$\alpha_{1,2,3}$ of $S^5$ at some values of the coordinates
$\theta$ and $\chi$.
The embedding implies that there are two world volume 
scalars $\theta$ and $\chi$. 
We choose the scalar $\chi$ to be trivial 
so that the D7-branes sit at a fixed value of $\chi$, but we take the scalar 
$\theta$, which is dual to the mass operator of the $\mathcal{N}=2$ theory,
as a function of $r$. 
We consider the diagonal $U(1)$ worldvolume gauge fields $A_\mu$, which are dual to 
$U(1)$ current $J^\mu$ of the dual field theory.
Recall that in the relativistic setup of Reference~\cite{Karch:2007pd},
the constant background electric field, $F_{ty}=-E$, was introduced.
To obtain the counterpart, we express this in the light-cone coordinates
(\ref{eq:bLC}) and get the corresponding gauge fields
\begin{align}\label{eq:DCgaueg}
	A_+ = {E_b }\, y + h_+ (r) \;, \qquad A_- = 2 \, b^2 \, {E_b}\, y + h_- (r)	 \;, 
	\qquad  A_y =  {A_y} (r) \ , 
\end{align}
where we have included the fluctuation fields that depend only on $r$ and
redefined the electric field $E_b := E/(2b)$, following the convention
of Reference~\cite{Ammon:2010eq}. 
Note that this gauge-field configuration is identical to the one adopted
in Reference~\cite{Ammon:2010eq}.%
\footnote{
 The sign difference in $A_-$ stems from the difference in the definition
 of light-cone coordinates.
}

Since we do not have NSNS $B$-field nor the coupling to RR-potentials,
the DBI action of the probes takes the form
\begin{equation}
	S_{D7} = - N_f T_{D7} \int d^8 \xi e^{-\Phi} \sqrt{-\det (g_{D7} + (2\pi\alpha') F)} \ , 
\end{equation}
where $T_{D7}, \xi$, $F$ and $\Phi$ are the D-brane tension, 
worldvolume coordinates, the $U(1)$ field strength and the dilaton, respectively.
The metric $g_{D7}$ is the pullback of the spacetime metric with respect to
the aforementioned embedding map.
Explicitly,
\begin{align}
	&ds_{D7}^2 = g_{++} \, dx^{+2} + 2 g_{+-} \,dx^+ dx^- + g_{--} \,dx^{-2} + g_{yy} \,d y^2 + g_{zz} \, d z^2  
	\nonumber \; \\ 
	&\qquad + g^{D7}_{rr} \, dr^2 + \cos^2 \theta \,(\sigma_1^2 +\sigma_2^2 + \sigma_3^2 ) \;,
\end{align}
where $g^{D7}_{rr} := g_{rr} + \theta' (r)^2$ and the prime denotes the derivative with respect to $r$. 
With our gauge fields, the action has the form   
\begin{equation}
	S_{D7} = - {\cal N} \int dr \sqrt{-\det M} \; , 
\end{equation}
where both sides are divided by the infinite volume of the field theory directions and we are working 
with the action density. We have defined ${\cal N} := 2\pi^2 N_f T_{D7}$
with $2\pi^2$ from the trivial integration of $S^3$ and%
\footnote{
 The contribution from the coordinates $\alpha_{2,3}$ in $G_{\alpha_2\alpha_3}$
 decouples from the others and becomes a multiplicative factor of the DBI action,
 while the factor $G_{\alpha_2\alpha_3}$ in the Schr\"dinger case \cite{Ammon:2010eq}
 couples with the $B_{\mu\nu}$ fields 
 and contributes to the conductivity calculation in a nontrivial way.
}
\begin{align}
	&\det M =  G_{\alpha_2\alpha_3} \, g_{\alpha_1\alpha_1}(r)\, g_{yy}(r) \,\Big[ g_{rr}^{D7}(r) 
	\left( \tilde E_b^2\, G_{3}+G_{+-} \,g_{yy}(r) \right) 
	+G_{+-} \tilde A_y'(r)^2  \nonumber \\ 
	&  \qquad \qquad  + g_{--}(r) \,g_{yy}(r) \left( \tilde h_+'(r) -\tilde  h_-'(r) \right)^2 
	+ \tilde E_b^2 \left( 2 b ^2 \,\tilde h_+'(r) - \tilde  h_-'(r) \right)^2  \Big] \ . 
\end{align}
Here the tildes indicate that the quantities are scaled with
the factor of $2\pi\alpha'$, such as $\tilde F = (2\pi\alpha') F$.
The sub-matrix determinants we need are
\begin{gather}
	G_{+-} = g_{++}(r) \,g_{--}(r) -g_{+-}(r)^2 \ , \qquad   
	G_{3} = 4 b ^4 \,g_{++}(r)-4 b ^2 \,g_{+-}(r) + g_{--}(r)  \ ,  \nonumber \\
 	G_{\alpha_2\alpha_3} = \frac{g_{\alpha_2\alpha_2}(r)\, g_{\alpha_3\alpha_3}(r) 
 	-g_{\alpha_2\alpha_3}(r)^2}{\sin^2 \alpha_1} \ .  
\end{gather}

The equations of motion of the gauge fields are constants of motion 
because the action consists of their derivatives only with respect to $r$.
Thus there are three constants of motion
\begin{align}
	&\langle J^+ \rangle := \frac{\delta {\cal L}}{\delta h_+'}
	= H \, 
	\left([ 4 \tilde E_b^2 \, b ^4+g_{--}(r)  \, g_{yy}(r) ] \tilde h_
	+'(r)- [2 \tilde E_b^2  \, b ^2+g_{+-}(r)  \, g_{yy}(r) ] \tilde h_-'(r)\right) \nonumber \\
	&\langle J^- \rangle := \frac{\delta {\cal L}}{\delta h_-'}
	=- H \,
	\left([ 2 \tilde E_b^2 \,  b ^2+g_{+-}(r)  \, g_{yy}(r) ] \tilde h_
	+'(r)- [\tilde E_b^2+g_{++}(r) \,  g_{yy}(r) ] \tilde h_-'(r)\right) \nonumber \\
	&\langle J^y \rangle  := \frac{\delta {\cal L}}{\delta A_y'}
	= H \,  G_{+-} \tilde A_y'(r) \; \\
	& \text{with} \quad H := \tilde {\cal N} \, \frac{G_{\alpha_2\alpha_3}
	g_{\alpha_1\alpha_1}(r) g_{yy}(r)}{\sqrt{-\det M}} \;, \nonumber 
\end{align}
where ${\cal L}$ is the Lagrangian of the DBI action. 
The quantities have their physical meanings as light-cone charge density $\langle J^+ \rangle$, 
light-cone current along $x^-$ direction $\langle J^- \rangle$, which is not 
independent but directly connected to $\langle J^+ \rangle$, and current along 
$y$ direction $\langle J^y \rangle$.

After solving the equations and plugging those solutions back into the action, 
we have the on-shell action  
\begin{equation}\label{OnShellActionDBI}
	S_{D7} = - 2\pi\alpha'\,{\cal N}^2 \int dr~ G_{\alpha_2\alpha_3}  \, g_{\alpha_1\alpha_1}(r) 
	\sqrt{g_{rr}^{D7}(r)}  \, g_{yy}(r) 
	\sqrt{\frac{\tilde E_b^2  \, G_{3}+G_{+-}  \, g_{yy}(r)}{U(r) - V(r) }} \;, 
\end{equation}
where 
\begin{align}
	&U(r) = - \frac{ \langle J^y \rangle ^2}{G_{+-}}-   \tilde {\cal N}^2  G_{\alpha_2\alpha_3} 
	g_{\alpha_1\alpha_1}(r)~g_{yy}(r)  \ ,  \\
	&V(r) = \frac{\tilde E_b^2 \left(\langle J^+ \rangle +2 \langle J^- 
	\rangle b ^2\right)^2+\left(\langle J^+ \rangle ^2 g_{++}(r)+\langle J^- 
	\rangle (2 \langle J^+ \rangle g_{+-}(r)+\langle J^- \rangle g_{--}(r))\right) 
	g_{yy}(r)}{g_{yy}(r) \left( \tilde E_b^2 G_{3}+G_{+-} g_{yy}(r)\right)} 
	\ .  \nonumber 
\end{align}

We concentrate on the last square root factor of the action (\ref{OnShellActionDBI}) 
and demand this to be real all the way from the horizon to the boundary \cite{Karch:2007pd}.
In the square root,
the numerator changes sign somewhere 
between the horizon and the boundary. This can be easily seen 
by the explicit expression of the numerator with the metric (\ref{eq:fullads5s5metricinconductivity}),
\beq
	\tilde E_b^2 G_{3}+G_{+-} g_{yy}(r) = \frac{r^2 
	\left(-r^4+ r_H^4+4 \tilde E_b^2 R^4  b ^2\right)}{R^6} \ .
\eeq 
At horizon $r=r_H$, this quantity is positive and it changes sign as $r$ is increased.
We assign the value of $r$ where 
this numerator changes sign as $r_*$ and we have 
\beq
\Big[ \tilde E_b^2 G_{3}+G_{+-} g_{yy}(r) \Big]_{r=r_*} = 0 \ . 
\label{rStarCondition}
\eeq

For the on-shell action (\ref{OnShellActionDBI}) to be real, 
the denominator should also vanish at $r=r_*$,
so we demand $U(r_*) - V(r_*) = 0$.
For this to happen, 
the numerator of the function $V$ should vanish at $r=r_*$ 
at least as fast as (\ref{rStarCondition}). 
Setting the numerator of $V$ to be zero at $r=r_*$, we get 
\beq
	\langle J^- \rangle = -\frac{ 2\tilde E_b^2  b ^2+g_{+-}(r) g_{yy}(r) }
	{4 \tilde E_b^2 b ^4+g_{--}(r) g_{yy}(r)} \bigg\vert_{r=r_*}  \langle J^+ \rangle \ ,
\eeq
which reduced to $\langle J^- \rangle = \langle J^+ \rangle / (2\beta^2)$ in the absence of 
the electric field. 

By plugging this condition in the equation $V(r_*)= U(r_*)$, 
we obtain the expression of the current along $y-$direction as 
\beq
	\langle J^y \rangle^2  = \frac{ \tilde E_b^2 G_3}{g_{yy}(r)} 
	\Big[ \tilde {\cal N}^2 G_{\alpha_2\alpha_3} g_{\alpha_1\alpha_1}(r) 
	 g_{yy}(r)+\frac{\langle J^+ \rangle^2}{4\tilde E_b^2  b ^4
	+g_{--}(r) g_{yy}(r)} \Big]  \bigg\vert_{r=r_*}  \nonumber \ .
\eeq
Using Ohm's law, we get
\begin{align}
    \sigma &= 
    2\pi\alpha' \sqrt{\frac{ G_3 }{g_{yy}(r_*)} }\left( \tilde {\cal N}^2 
    G_{\alpha_2\alpha_3} g_{\alpha_1\alpha_1}(r_*)  g_{yy}(r_*)
	+\frac{\langle J^+ \rangle^2}{4\tilde E_b^2  b ^4
	+g_{--}(r_*) g_{yy}(r_*)} \right)^{\frac{1}{2}}  \nonumber \\
	&=2\pi\alpha'\sqrt{\frac{\tilde  {\cal N}^2 b ^2 \cos^6 \theta (r_*)}{16 } 
	\sqrt{4 \tilde E_b^2 b ^2+ R^4 \pi ^4 T^4 b ^4} 
	+ \frac{4 \langle J^+ \rangle^2 }{4 \tilde E_b^2 b ^2+R^4 \pi ^4 T^4 b ^4}} 
	\label{conductivityformula}
\end{align}
where we used the relation $r_H = R^2 \pi T b $. 
This is the final expression for conductivity and is the analogue of 
the equation (3.27) of Reference~\cite{Ammon:2010eq}.
In the reference,
the origin of the first term was identified as
the Schwinger pair production.
Also notice that this is a dimensionless quantity,
which is appropriate for the conductivity in $2+1$ dimensional field theory.  

Our result and that of Reference~\cite{Ammon:2010eq} are different.
However, this is {\it not} inconsistent,
because while we have derived the conductivity
which is exactly the counterpart of Karch-O'Bannon \cite{Karch:2007pd},
the setup of Reference~\cite{Ammon:2010eq} may not be so.
In fact, we have checked that the null Melvin twist on the system
of Karch and O'Bannon (with the constant background electric field)
yields very different supergravity solution from the setup of Reference~\cite{Ammon:2010eq}.
Therefore, it appears that despite the same gauge field configuration
in Equation~(\ref{eq:DCgaueg}), the DC conductivities belong to
different non-relativistic systems. 
We, however, demonstrate in the rest of this section
that they become identical in a few important limiting cases.

Let us take the limit $\tilde E_b \ll b(RT)^2$,
for weak electric fields compared to temperature.
The conductivity becomes
\beq
	\sigma \approx 
        2\pi\alpha'\sqrt{\frac{4 \langle J^+ \rangle^2 }{\pi^4 b^4 (RT)^4}
	+\frac{\tilde {\cal N}^2  \cos^6 \theta (r_*)}{16 }  \pi^2b^4(RT)^2}  \;.
	\label{smallelectyricfieldlimit}
\eeq
Note that there is a difference in the $\cos \theta$ factor compared to 
Reference~\cite{Ammon:2010eq}.
For $\cos \theta (r_*) \approx 0$ or at low temperature, the second term of 
(\ref{smallelectyricfieldlimit}) is suppressed and the conductivity is 
\beq
	\sigma \approx 
        2\pi\alpha' \frac{2 \langle J^+ \rangle }{\pi^2b^2(RT)^2} \;,
	\label{smallTsmallelectyricfieldlimit}
\eeq
and this is identical to the result of Reference~\cite{Ammon:2010eq}.

For the opposite limit $\tilde E_b \gg b(RT)^2$, we have
\beq
	\sigma
	\approx \, 2\pi\alpha' \sqrt{\frac{\langle J^+ \rangle^2 }{ b^2 \tilde E_b^2}
	+\frac{\tilde  {\cal N}^2  \cos^6 \theta (r_*)}{8}  \, b^3\tilde E_b} \ . 
	\label{ElectricFieldBigLimit}
\eeq
In this limit, the conductivity is identical to the (3.30) in \cite{Ammon:2010eq} 
without further assumptions to the field $\theta$.
If we further take a limit with a small density $\langle J^+ \rangle \approx 0$ and 
small $\theta (r_*)$, the conductivity, $\sigma \approx (2\pi \alpha')\, 
\sqrt{\frac{ \tilde {\cal N}^2 }{8} \tilde E_b b^3 }$, is mainly from the Schwinger pair 
production. On the other hand, in the opposite limit with a large density 
$\langle J^+ \rangle \approx 0$ and $\theta (r_*) \approx \pi/2$ 
in the equation (\ref{ElectricFieldBigLimit}), we get $\sigma \approx (2\pi \alpha')\, 
\frac{\langle J^+ \rangle }{\tilde E_b \, b }$, which is identical to the previous 
result in the same limit \cite{Ammon:2010eq}.

Thus we see that our result and that of Reference~\cite{Ammon:2010eq} are
identical in some limiting cases, especially when the limits bring the
expressions independent of the scalar profile $\theta(r)$.
This is remarkable considering the drastic differences in the backgrounds,
where one of them even involves non-vanishing NSNS $B$-field.

%%%%%%%%%%%%%%%%%%%%%%%%%%%%%%%%%%%%%%%%%%%%%%%%%%%%%%%%%%%%%%%%%%%%%
\section{Concluding Remarks}\label{sec:conclude}
%%%%%%%%%%%%%%%%%%%%%%%%%%%%%%%%%%%%%%%%%%%%%%%%%%%%%%%%%%%%%%%%%%%%%
In this paper, we have taken less discussed approach to the non-relativistic
generalization of the AdS/CFT, essentially by following
References~\cite{Goldberger:2008vg,Barbon:2008bg,Maldacena:2008wh}.
The procedure is to simply adopt special light-cone coordinates
and re-interpret one of them as the non-relativistic time.
We have shown that various black hole properties can be obtained
in the same way as the relativistic counterparts, and our results
are consistent with the ones directly derived from the Schr\"odinger
black holes.
We hope that we have demonstrated and conveyed the simplicity and well-defined
nature of the procedure.
In particular, we have emphasized that the standard boundary of the geometry
allowed us to utilize the usual Brown-York procedure to obtain the thermodynamic
quantities of non-relativistic systems.
We also derived the scalar two-point correlator and the associated shear
viscosity, as well as the DC conductivity which is the non-relativistic
counterpart of Reference~\cite{Karch:2007pd}.
All those computations are no more complicated than the relativistic
counterparts.

While we think that our procedure is very powerful in dealing with the
geometry with Schr\"odinger symmetry, it is unclear how to apply
or generalize the procedure to different non-relativistic systems,
such as Lifshitz spacetime.
Schr\"odinger and Lifshitz systems are special cases of
scale invariant non-relativistic systems, and
they are characterized by the relative
scaling factor of time and space, known as the dynamical exponent.
One way to incorporate the dynamical exponent for the Schr\"odinger space 
is demonstrated explicitly using the parameter $b$ \cite{Kim:2010}.

The calculation of DC conductivity is carried out with the gauge fields 
that is corresponding to the electric field, $F_{ty}$,
in the relativistic AdS space. 
This is not really natural from the point of view of
the non-relativistic theory, because as we have seen, it gives rise to
nonzero $F_{-y}$ as well as $F_{+y}$.
The reason we take this particular form of gauge fields is to
compare directly to the known results.
It is more appropriate to have background field only along 
the $F_{+y}$ to calculate the DC conductivity, 
which seems to give us more interesting DC conductivity results,
while the same gauge field does not give any instability to
the Schr\"odinger space. 
This is discussed recently in \cite{Kim:2010}.

It is of a great interest to figure out the similarities and differences 
between the two different geometric realizations of the Schr\"odinger 
geometry, Schr\"odinger background and AdS in light-cone, in detail.

%%%%%%%%%%%%%%%%%%%%%%%%%%%%%%%%%%%%%%%%%%%%%%%%%%%%%%%%%%%%%%%%%%%%%%%%
\section*{Acknowledgments}
%%%%%%%%%%%%%%%%%%%%%%%%%%%%%%%%%%%%%%%%%%%%%%%%%%%%%%%%%%%%%%%%%%%%%%%%
We would like to thank to K. Bardakci, O. Ganor, S. Hartnoll, J. Hartong, P. Ho\v{r}ava, 
E. Kiritsis, M. Lippert, A. O'Bannon, C. Panagopoulos and M. Taylor for useful
conversations and correspondence. 
BSK also thanks to  S-J. Hyun, Y-S. Myung, S-J. Sin, P-j. Yi 
for discussions and comments during his visits to CQUeST and KIAS, Seoul.

BSK is supported through MEXT-CT-2006-039047 and also partially supported by  
a European Union grant FP7-REGPOT-2008-1-CreteHEP Cosmo-228644 and 
by ANR grant  STR-COSMO, ANR-09-BLAN-0157.
DY is supported by Marie Curie International Incoming Fellowship
FP7-PEOPLE-IIF-2008.

\bigskip
\bigskip

%%%%%%%%%%%%%%%%%%%%%%%%%%%%%%%%%%%%%%%%%%%%%%%%%%%%%%%%%%%%%%%%%%%%%
%%%%%%%%%%%%%%%%%%%%%%%%%%%%%%%%%%%%%%%%%%%%%%%%%%%%%%%%%%%%%%%%%%%%%
\appendix
%%%%%%%%%%%%%%%%%%%%%%%%%%%%%%%%%%%%%%%%%%%%%%%%%%%%%%%%%%%%%%%%%%%%%
%%%%%%%%%%%%%%%%%%%%%%%%%%%%%%%%%%%%%%%%%%%%%%%%%%%%%%%%%%%%%%%%%%%%%

%%%%%%%%%%%%%%%%%%%%%%%%%%%%%%%%%%%%%%%%%%%%%%%%%%%%%%%%%%%%%%%%%%%%%
\section{Infinite Boost}\label{app:InfiniteB}
%%%%%%%%%%%%%%%%%%%%%%%%%%%%%%%%%%%%%%%%%%%%%%%%%%%%%%%%%%%%%%%%%%%%%
In this appendix, we demonstrate that the quantities of Schr\"odinger
black holes can be extracted from the AdS black hole counterparts in
infinite momentum frame.
We reproduce the results of Section~\ref{sec:procedure}
as a prototype example of this technique.
The motivation comes from Susskind's work in 1967
\cite{Susskind:1967rg}, where he shows that a theory (a system of
free scalar bosons) on a infinitely boosted frame resembles a non-relativistic
system.
We have a gravitational system but the procedure is somewhat similar
to Susskind's.

We start from the metric (\ref{eq:pOriginal}) and apply the Lorentz
transformation
\begin{align}
  \begin{pmatrix} t \\ x \end{pmatrix}
  =
  \begin{pmatrix}
    \cosh\zeta & \sinh\zeta \\
    \sinh\zeta & \cosh\zeta
  \end{pmatrix}
  \begin{pmatrix} t' \\ x' \end{pmatrix}
  \;.
\end{align}
Then the metric becomes
\begin{align}
  ds_5^2 =& \bigg(\frac{r}{R}\bigg)^2 \big\{ (\sinh^2\zeta-h\cosh^2\zeta)dt^2
         + 2(1-h)\sinh\zeta\cosh\zeta dtdx 
  \nonumber\\         
         &+ (\cosh^2\zeta-h\sinh^2\zeta)dx^2 + dy^2 + dz^2 \big\}
         + \bigg(\frac{R}{r}\bigg)^2 h^{-1} dr^2
  \;,
\end{align}
where we have omitted the primes on the coordinates $t$ and $x$.
The ADM form of the metric is
\begin{align}
  ds_5^2 =& \bigg(\frac{r}{R}\bigg)^2 \bigg[ -\frac{hdt^2}{\cosh^2\zeta-h\sinh^2\zeta} 
         + (\cosh^2\zeta-h\sinh^2\zeta) \bigg\{ dx + 
         \frac{(1-h)\sinh\zeta\cosh\zeta}{\cosh^2\zeta-h\sinh^2\zeta}dt \bigg\}^2 
  \nonumber\\         
         &+dy^2+dz^2 \bigg] + \bigg(\frac{R}{r}\bigg)^2 h^{-1} dr^2
  \;.
\end{align}
From this, one obtains the entropy, horizon velocity and temperature
in terms of the rapidity $\zeta$.
Moreover, the mass and the momentum in $x$-direction can be computed
through the Brown-York procedure as detailed in Section~\ref{sec:procedure}.
The result is
\begin{align}
  S =& \frac{V_3}{4G_5} \bigg(\frac{r_H}{R}\bigg)^3 \cosh\zeta
    = \frac{V_3}{8G_5} \bigg(\frac{r_H}{R}\bigg)^3 \bigg( \frac{1}{\epsilon}+\epsilon \bigg)
  \nonumber\\
  \Omega_H =& -\tanh\zeta
           = -1+2\epsilon^2-2\epsilon^4+\mathcal{O}(\epsilon^6)
  \nonumber\\
  \beta =& \frac{\pi R^2}{r_H} \cosh\zeta
    = \frac{\pi R^2}{2r_H} \bigg( \frac{1}{\epsilon}+\epsilon \bigg)
  \nonumber\\
  M =& \frac{3r_H^4V_3}{16\pi G_5 R^5} \frac{4\cosh^2\zeta-1}{3}
    = \frac{r_H^4V_3}{16\pi G_5 R^5} \bigg( \frac{1}{\epsilon^2}+1+\epsilon^2 \bigg)
  \nonumber\\
  J_x =& -V_3\frac{r_H^4\sinh\zeta\cosh\zeta}{4\pi G_5R^5}
    = -\frac{r_H^4V_3}{16\pi G_5R^5} \bigg( \frac{1}{\epsilon^2} - \epsilon^2 \bigg)
  \;,
\end{align}
where we have defined $\epsilon := e^{-\zeta}$ and we are interested
in the limit $\epsilon\to 0$.

As a non-relativistic system, we pick the most singular terms in $S$,
$\beta$ and $J_x$.
This procedure is similar to the scaling done by Susskind
\cite{Susskind:1967rg}.
For the horizon velocity $\Omega_H$, we drop the speed of light ($-1$)
and take the next term in the $\epsilon$ expansion.
This is because there is an infinite contribution to the mass from
the momentum in $x$-direction (the most singular term in $\Omega_HJ_x$)
and following Reference~\cite{Susskind:1967rg}, we chose to drop this
infinite contribution.
Consequently, we must choose the finite term in the quantity $M$.
After the replacement $\epsilon\to 1/2b$, we find that the extracted
non-relativistic quantities are identical to Equation~(\ref{eq:pThermo}).

%%%%%%%%%%%%%%%%%%%%%%%%%%%%%%%%%%%%%%%%%%%%%%%%%%%%%%%%%%%%%%%%%%%%%
\section{Geometric Quantities in Non-Coordinate
Frames}\label{app:nonCoords}
%%%%%%%%%%%%%%%%%%%%%%%%%%%%%%%%%%%%%%%%%%%%%%%%%%%%%%%%%%%%%%%%%%%%%
We summarize the frames adopted for the spherical ($k=+1$) and
hyperbolic ($k=-1$) black holes and remind the reader of the geometric
quantities in a non-coordinate basis.

%%%%%%%%%%%%%%%%%%%%%%%%%%%%%%%%%%%%%%%%%%%%%%%%%%%%%%%%%%%%%%%%%%%%%
\subsection{Frame for $k=+1$}
In Section~\ref{sec:RCharged}, following coframe is introduced;
\begin{align}
  \omega^+ &= b(dt+\eta_{+1})
  \;,\qquad
  \omega^- = \frac{1}{2b}(dt-\eta_{+1})
  \;,
  \nonumber\\
  \omega^\theta =& \frac{R}{2} d\theta
  \;,\qquad
  \omega^\phi = \frac{R}{2} \sin\theta d\phi
  \;,\qquad
  \omega^r = dr
  \;,
\end{align}
where $\eta_{+1}:=(R/2)(d\psi+\cos\theta d\phi)$.
The dual frame $\{e_\mu\}$ that satisfy $\omega^\mu e_\nu = \delta^\mu_\nu$ is
\begin{align}
  &\quad
  e_+ = \frac{1}{2b}(\partial_t + \frac{2}{R}\partial_\psi)
  \;,\qquad
  e_- = b(\partial_t - \frac{2}{R}\partial_\psi)
  \;,
  \nonumber\\
  e_\theta =& \frac{2}{R} \partial_\theta
  \;,\qquad
  e_\phi = \frac{2}{R} \big( -\cot\theta\partial_\psi
        + \frac{1}{\sin\theta}\partial_\phi \big)
  \;,\qquad
  e_r = \partial_r
  \;.
\end{align}
This is a non-coordinate basis and we have the non-vanishing
Lie bracket
\begin{align}
  [e_\theta , e_\phi] = \frac{2b}{R}\,e_+ - \frac{1}{bR}e_-
  	-\frac{2}{R}\cot\theta\, e_\phi
  \;.
\end{align}

%%%%%%%%%%%%%%%%%%%%%%%%%%%%%%%%%%%%%%%%%%%%%%%%%%%%%%%%%%%%%%%%%%%%%
\subsection{Frame for $k=-1$}
For this case, the coframe introduced is
\begin{align}
  \omega^+ &= b(dt+\eta_{-1})
  \;,\qquad
  \omega^- = \frac{1}{2b}(dt-\eta_{-1})
  \;,
  \nonumber\\
  \omega^\chi =& \frac{R}{2} d\chi
  \;,\qquad
  \omega^\phi = \frac{R}{2} \sinh\chi d\phi
  \;,\qquad
  \omega^r = dr
  \;,
\end{align}
where $\eta_{-1}:=(R/2)(d\psi+\cosh\chi d\phi)$.
The dual frame is
\begin{align}
  &\quad
  e_+ = \frac{1}{2b}(\partial_t + \frac{2}{R}\partial_\psi)
  \;,\qquad
  e_- = b(\partial_t - \frac{2}{R}\partial_\psi)
  \;,
  \nonumber\\
  e_\chi =& \frac{2}{R} \partial_\chi
  \;,\qquad
  e_\phi = \frac{2}{R} \big( -\coth\chi\partial_\psi
        + \frac{1}{\sinh\chi}\partial_\phi \big)
  \;,\qquad
  e_r = \partial_r
  \;.
\end{align}
The non-vanishing Lie bracket is
\begin{equation}
  [e_\chi , e_\phi] = -\frac{2b}{R}\,e_+ + \frac{1}{bR}e_-
  	-\frac{2}{R}\coth\chi\, e_\phi
  \;.
\end{equation}

%%%%%%%%%%%%%%%%%%%%%%%%%%%%%%%%%%%%%%%%%%%%%%%%%%%%%%%%%%%%%%%%%%%%%
\subsection{Geometric Quantities}
The geometric quantities are defined without referring to a frame,
especially in mathematical literature.
Coordinate expressions are popular among physicists but the usual
formulas must be modified when a non-coordinate basis is adopted.
The formulas below are taken from MTW~\cite{MTW}.

Given a frame whose Lie algebra product is
\begin{align}
  [e_\alpha,e_\beta] = {c_{\alpha\beta}}^\gamma e_\gamma
  \;,
\end{align}
the connection coefficients are
\begin{align}
  \Gamma_{\alpha\beta\gamma} = \frac{1}{2} (g_{\alpha\beta,\gamma}
  + g_{\alpha\gamma,\beta} - g_{\beta\gamma,\alpha}
  + c_{\alpha\beta\gamma} + c_{\alpha\gamma\beta} - c_{\beta\gamma\alpha})
  \;,
\end{align}
where the comma ``$,\mu$'' implies the derivation with respect to $e_\mu$.
Notice that the usual symmetry exchanging the indices $\beta$ and
$\gamma$ is not necessarily true.
The covariant derivative is given with respect to this connection.
So for instance, $F_{\mu\nu}:=A_{\nu;\mu}-A_{\mu;\nu}$, and since the
exchanging symmetry is lost, the semicolons here cannot be replaced
by just colons.
Also with the covariant derivative, the definition of the second fundamental
form (\ref{eq:2ndFF}) is not modified.

The Riemann tensor is
\begin{align}
  {R^\alpha}_{\beta\gamma\delta} = {\Gamma^\alpha}_{\beta\delta,\gamma}
  - {\Gamma^\alpha}_{\beta\gamma,\delta}
  + {\Gamma^\alpha}_{\mu\gamma}{\Gamma^\mu}_{\beta\delta}
  - {\Gamma^\alpha}_{\mu\delta}{\Gamma^\mu}_{\beta\gamma}
  - {\Gamma^\alpha}_{\beta\mu}{c_{\gamma\delta}}^\mu
  \;.
\end{align}
The Ricci tensor and the scalar curvature are given from this
Riemann tensor.

%%%%%%%%%%%%%%%%%%%%%%%%%%%%%%%%%%%%%%%%%%%%%%%%%%%%%%%%%%%%%%%%%%%%%
\section{Scalar Two-Point Correlator
 and Viscosity-Entropy Ratio}\label{app:ratio}
%%%%%%%%%%%%%%%%%%%%%%%%%%%%%%%%%%%%%%%%%%%%%%%%%%%%%%%%%%%%%%%%%%%%%

Using the metric (\ref{eq:planar}) with the time coordinate $x^+$,
we can exactly follow the procedure of
Son and Starinets \cite{Son:2002sd} to compute two-point correlators.
In particular, we need the correlator of the stress tensor in $yz$
components to obtain the viscosity.
However, as shown in Reference~\cite{Policastro:2002se}, this
is equivalent to the correlator of a minimally coupled scalar 
in the background.
Therefore, we sketch the calculation of the scalar two point
correlator here.

First, we introduce a new coordinate in the metric (\ref{eq:planar}),
according to
\begin{align}
  u := r_H^2/r^2
  \;.
\end{align}
Then the action for the scalar field becomes
\begin{align}
  &K \int d^4x \int_{0}^{1}du \sqrt{-g} \bigg[
      g^{uu}(\partial_u\phi)^2 + g^{\mu\nu}(\partial_\mu\phi)(\partial_\nu\phi)
    + m^2\phi^2  \bigg]
  \nonumber\\
  &= (KR^3) \bigg(\frac{r_H^2}{2R^4}\bigg) \int d^4x \int_{0}^{1}du
    u^{-2} \bigg[ \frac{4r_H^2}{R^4}hu(\partial_u\phi)^2 + (\partial_i\phi)^2
    + b^2(1-h^{-1})(\partial_+\phi)^2
    \nonumber\\
    &-(1+h^{-1})(\partial_+\phi)(\partial_-\phi)
    + \frac{1}{4b^2}(1-h^{-1})(\partial_-\phi)^2
     + m^2\bigg(\frac{r_H}{R}\bigg)^2u^{-1}\phi^2  \bigg]
  \;,
\end{align}
where we have $h=1-u^2$ and through the AdS/CFT dictionary,
$KR^3=-N_c^2/(16\pi^2)$.
After Fourier decomposing as
\begin{align}\label{eq:fourier}
  \phi(u,x^+,x^-,\vec y) = \int\frac{d^4k}{(2\pi)^4}
  e^{-i\omega x^+ + ik_-x^- + i\vec k \cdot \vec y} f_k (u) \phi_0(k)
  \;,
\end{align}
the equations of motion read
\begin{align}
  &f_k'' + (\ln[u^{-1}h])'f_k'
  \nonumber\\
  &-\frac{1}{uh}\bigg\{ q_i^2 + (1-h^{-1})w^2
  + 2(1+h^{-1}) w q_-
  + (1-h^{-1})q_-^2
  + \bigg(\frac{mR}{2}\bigg)^2u^{-1} \bigg\}f_k
  =0
  \;,
\end{align}
where the primes denote the derivatives with respect to
the variable $u$
and we have defined
\begin{align}
  w := b\frac{R^2}{2r_H}\omega
  \;,\quad
  q_- := \frac{1}{2b}\frac{R^2}{2r_H}k_-
  \quad\text{and}\quad
  q_i := \frac{R^2}{2r_H}k_i
  \;.
\end{align}
The differential equation is singular at the horizon $u=1$
and the idea is to extract the regular part of the function
and solve for it.
For this purpose, we set
\begin{align}
  f_k(u) = (1-u)^\nu F_k(u)
  \;,
\end{align}
and plug this into the differential equation.
For the massless case $m = 0$,
the regularity for the function $F_k(u)$ near $u=1$ determines that
\begin{align}
  \nu = \pm \frac{i}{2}\big(w-q_-\big)
  \;.
\end{align}
Since we are interested in the limit $q_-\to 0$,
we pick the negative sign for the incoming wave solution.
With this value of $\nu$ (and with $m=0$), the differential equation
for the regular part $F_k(u)$ is
\begin{align}
  F_k'' +& \bigg\{ -\frac{1+u^2}{u(1-u^2)} 
  + \frac{i( w - q_-)}{1-u} \bigg\}F_k'
  \nonumber\\
  &- \bigg\{ \frac{( w - q_-)^2}{4(1+u)^2}
    + \frac{2q_i^2 + 8 q_- w + i( w - q_-)}{2u(1-u^2)}
  \bigg\} F_k
  =0
  \;.
\end{align}
By assuming $w,q\ll 1$, we can solve this differential equation order
by order in those parameters, and the solution is
\begin{align}
  F_k(u) = 1 - \bigg\{ \frac{i}{2}(w-q_-)+q_i^2 \bigg\}\ln\frac{1+u}{2} 
  + \text{higher orders}
  \;,
\end{align}
where the regularity of the solution at $u=1$ and the boundary condition
$F_k(1)=1$ were imposed, by following Reference~\cite{Son:2002sd}.

To obtain the viscosity through Kubo's formula, we set $q_-=0=q_i$ and
write the solution as
\begin{align}
  f_k(u) = (1-u^2)^{-iw/2} + \text{higher orders}
  \;,
\end{align}
where the factor $2^{iw/2}$ was extracted from $\phi_0(k)$
in Equation~(\ref{eq:fourier}), in order
to satisfy the required boundary condition $f_k(0)=1$.%
\footnote{
  We could have avoided this awkward extraction of the factor $2^{iw/2}$
  by changing the boundary condition $F_k(1)=1$, but we chose to have
  the congruence with Son and Starinets \cite{Son:2002sd}.
}
Then the Son-Starinets prescription yields the retarded Green's function
\begin{align}
  G^R = ib \frac{N_c^2}{8\pi^2} \frac{r_H^3}{R^6} \omega
  \;.
\end{align}
This is the same result as Reference~\cite{Son:2002sd}, except for
the parameter $b$.
However, since the same parameter also appears in the entropy
(\ref{eq:pThermo}), the viscosity-entropy ratio is identical
to the relativistic case.

\pagebreak

%%%%%%%%%%%%%%%%%%%%%%%%%%%%%%%%%%%%%%%%%%%%%%%%%%%%%%%%%%%%%%%%%%%%%

\end{document}